\documentclass[12pt]{article}
\usepackage{graphicx,epsfig}
\usepackage{amssymb}
\usepackage{amsmath}

\newcommand{\bea}{\begin{eqnarray}}
\newcommand{\beq}{\begin{equation}}
\newcommand{\eea}{\end{eqnarray}}
\newcommand{\eeq}{\end{equation}}

\newcommand{\barr}{\begin{array}}
\newcommand{\earr}{\end{array}}

\newcommand{\lsim}{\raise0.3ex\hbox{$\;<$\kern-0.75em\raise-1.1ex\hbox{$\sim\;$}}}
\newcommand{\gsim}{\raise0.3ex\hbox{$\;>$\kern-0.75em\raise-1.1ex\hbox{$\sim\;$}}}

\newcommand{\eq}[1]{Eq.~(\ref{#1})}

\newcommand{\bpm}{\begin{pmatrix}}
\newcommand{\epm}{\end{pmatrix}}

\begin{document}

\begin{center}
\begin{flushright}
CERN-PH-TH/2004-108\\
MCTP-04-33\\
OUTP-0412P\\
SHEP 04/17\\
\end{flushright}

{\Large \textbf{\ K\"{a}hler corrections and softly broken family symmetries}%
}\\[0pt]
\vspace*{0.5cm} S.\ F.\ King{\footnote{%
sfk$@$hep.phys.soton.ac.uk}} and I.\ N.\ R.\ Peddie{\footnote{%
i.peddie$@$hep.phys.soton.ac.uk}} \\[0pt]
{\small School of Physics and Astronomy, U. of Southampton, Southampton SO17
1BJ, UK}\\[0.2cm]
Graham\ G.\ Ross{\footnote{%
g.ross1$@$physics.ox.ac.uk}} \\[0pt]
{\small The Rudolf Peierls Centre for Theoretical Physics, 1 Keble Road, }\\[%
0pt]
{\small Oxford, OX1 3NP} \\[0.2cm]
Liliana Velasco-Sevilla{\footnote{%
lvelsev$@$umich.edu}} \\[0pt]
{\small Michigan Center for Theoretical Physics, Randall Laboratory, U. of
Michigan,}\\[0pt]
{\small 500 E University Ave., Ann Arbor, MI 48109, USA} \\[0.2cm]
Oscar Vives{\footnote{%
oscar.vives$@$cern.ch}} \\[0pt]
{\small Theory Division, CERN, CH-1211, Geneva 23, Switzerland\bigskip }

\textbf{Abstract}
\end{center}

{\small Spontaneously broken family symmetry provides a promising origin for
the observed quark and lepton mass and mixing angle structure. In a
supersymmetric theory such structure comes from a combination of the
contributions from the superpotential and the K\"{a}hler potential. The
superpotential effects have been widely studied but relatively little
attention has been given to the effects of the K\"{a}hler sector. In this
paper we develop techniques to simplify the analysis of such K\"{a}hler
effects. Using them we show that in the class of theories with an
hierarchical structure for the Yukawa couplings the K\"{a}hler corrections
to both the masses and mixing angles are subdominant. This is true even in
cases that texture zeros are filled in by the terms coming from the K\"{a}%
hler potential.}

\section{Introduction}

\label{sec:intro}

The origin of the structure of the fermion Yukawa couplings is one of the
most intriguing puzzles left unanswered by the Standard Model. The
hierarchical pattern of fermion masses and quark mixing angles strongly
suggests the existence of a spontaneously broken family symmetry with the
order parameter of breaking (the vacuum expectation value (vev) of one or
more scalar familon fields) providing the small expansion parameter(s). This
has been the most popular strategy to try to improve our understanding of
the flavour structures in nature. In this scheme the (usually
Supersymmetric) Standard Model is extended by a gauge or global family
symmetry $G_{F}$ which is then spontaneously broken. The Yukawa couplings
(or just those associated with the two lighter generations) are not allowed
in the limit of unbroken family symmetry but are filled in by higher
dimension operators involving powers of the familon field(s). Thus below the
scale of $G_{F}$ breaking, we have an effective theory resembling the
Supersymmetric Standard Model where the Yukawa couplings (with the possible
exception of the third family) and all the different flavour structures are
given by non-renormalisable operators in the superpotential of the kind, 
\begin{equation}
\psi \psi^{c}H\left( \frac{\displaystyle{\langle \theta \rangle }}{%
\displaystyle{M}}\right) ^{n}  \label{one}
\end{equation}%
where $\psi$ and $\psi^c$ denote quark/lepton superfields, $H$ is a Higgs
superfield, $M$ is the heavy messenger mass corresponding to the
intermediate state in the Froggatt-Nielsen mechanism and $\langle \theta
\rangle $ is the familon vev that breaks $G_{F}$ such that $\langle \theta
\rangle /M$ is a small expansion parameter \cite{froggatt}. In the
following, we do not specify the family group or the exact mechanism of
symmetry breaking as our conclusions are equally applicable to (Abelian or
non-Abelian) flavour theories generating a given structure in the Yukawa 
couplings \cite{flavth}.

The generation of Yukawa couplings (or other holomorphic couplings in
the superpotential) though non-renormalisable operators is not the
only effect of integrating out the heavy fields in the low energy
effective theory. It is well known that the non-holomorphic couplings
involving the kinetic terms and gauge couplings also receive
corrections from the flavour breaking terms. This implies a non
canonical K\"{a}hler potential, different from the identity in flavour
space. In determining the physical implications of the theory it is
much simpler to work in a theory with canonical kinetic terms and this
can be done by choosing a linear combination of the chiral superfields
such that the new fields have a canonical K\"{a}hler potential
\cite{weinberg}. As originally shown by Leurer, Nir and Seiberg  
\cite{leurernir} 
this transformation in the chiral superfields
consists of a rotation in flavour space and a rescaling of the
fields. 
However, even after this field redefinition, we can still
perform further arbitrary unitary rotations of the chiral superfields
which will preserve the canonical form of the K\"{a}hler
potential. Clearly any superfield field redefinitions in the
K\"{a}hler potential must be performed consistently for all the
superfields in the theory and this will result in a transformation of
the superpotential couplings when written in terms of the new chiral
superfields. This transformation of the Yukawa couplings is the main
subject of this work and we are especially interested in the
observable effects of this transformation on the physical masses and
mixing angles. In fact, in the literature it is often stated that
these field redefinitions can have very important observable effects
in quark and squark mixings \cite{dudas,dreiner,peddie}. 
That this is not the case in specific models has been stressed in
\cite{leurernir,timjones}. Here we generalise this result and show that, 
at least for the case of an
hierarchical Yukawa textures for the up and down
sectors, the effect of the K\"{a}hler potential, is always
sub-dominant and cannot change the structure coming from the
superpotential.  In the presence of a hierarchical texture ordered 
by an underlying family symmetry, the
structure of the K\"ahler potential is such that the off-diagonal
elements are given by powers of the same small expansion parameter
that generates the hierarchy in the Yukawa matrices. Under these
conditions we show that we can choose an upper triangular form for the
inverse of the square root of the K\"ahler metric which brings the
fields to the canonical basis. Using this form we can prove the
subdominance of the K\"ahler corrections to the Yukawa matrices. Even
in cases without a clear hierarchy we show that only unknown coefficients ${\mathcal{O}} (1)$
can be changed without modifying the structure of the observable
mixings and masses, consistent with the results of \cite{leurernir}. 
Therefore our conclusions apply to all 
flavour models with hierarchical Yukawa textures considered in the literature.

\section{The K\"{a}hler metric}

After the flavour symmetry is spontaneously broken we obtain a certain
Yukawa texture given by non-renormalisable operators which are functions of
the flavon vevs as in Eq.~(\ref{one}). In the same way the effective K\"{a}%
hler potential will be a general non-renormalisable real function invariant
under all the symmetries of the theory coupling the superfield combinations $%
\psi _{i}^{\dagger }\psi _{j}$ to the flavon fields, and similarly for ${%
\psi^c}_{i}^{\dagger }\psi^c_{j}$, where $i,j$ are flavour indices. The
terms $\psi _{i}^{\dagger }\psi _{i}$, ${\psi^c}_{i}^{\dagger }\psi^c_{i}$
without flavon superfields are clearly invariant under gauge, flavour and
global symmetries and hence give rise to a family universal contribution.
However, family symmetry breaking terms involving flavon superfields give
rise to important corrections \cite{softflavour,alignment,leurernir}. In
fact, it is interesting to notice that, due to the non-holomorphicity of the
K\"{a}hler potential, new terms are allowed with different structure from
the terms that appear in the Yukawa couplings of the superpotential.

In general the matter fields do not have canonical wave functions (kinetic
terms) in the symmetry eigenstate basis $\hat{\psi}_{i}^{c}$, $\hat{\psi}%
_{j} $ \cite{peddie}. Rather, flavon field vevs contribute to the diagonal
terms and also generate new flavour off-diagonal entries. Thus, we have now
non-canonical kinetic terms and we must redefine the fields to obtain
canonical kinetic terms. The effect of these redefinitions, which can be
regarded as wave function corrections, on the Yukawa couplings and other
couplings in the theory may be determined after this field redefinition, $%
\hat{\psi}=N\psi $.

To obtain canonical kinetic terms we have to redefine the fields to go to
the canonical basis by the inverse of the square root of the K\"{a}hler
metric $K$ given by 
\begin{equation}
\hat{\psi}^{\dagger }K\hat{\psi}=(N\psi )^{\dagger }(N^{-1})^{\dagger
}N^{-1}N\psi  \label{kahlerdec}
\end{equation}%
Thus $K=(N^{-1})^{\dagger }N^{-1}$ and hence $N=K^{-1/2}$, as claimed above.
Using Supergravity (SUGRA) equations, the K\"{a}hler\ metric is obtained as $%
K_{\bar{a}b}=\partial ^{2}G/(\partial \Phi _{a}^{\dagger }\partial \Phi
^{b}) $ with $G$ the K\"{a}hler\ function and it determines both the Kinetic
terms and the non-canonically normalised soft scalar mass squared matrices $%
\hat{m}_{\bar{a}b}^{2}$. In SUGRA, where $K_{\bar{a}b}$ represents a metric, 
$N^{-1} $ is also a Hermitian matrix, such that $N^{-1}=(N^{-1})^{\dagger }$
and hence it can be conventionally written as \cite{dreiner,peddie} 
\begin{equation*}
K=(N^{-1})^{\dagger }N^{-1}=V^{\dagger }X^{2}V\Leftrightarrow
N^{-1}=V^{\dagger }XV
\end{equation*}%
with V a unitary matrix diagonalising the Hermitian matrix $K$ and $X$ the
square root of the eigenvalues of $K$. We call this solution the
\textquotedblleft standard\textquotedblright\ form of $N^{-1}$. Note that if 
$N^{-1}$ is a solution of Eq.~(\ref{kahlerdec}) then also $R.N^{-1}$ is a
solution of Eq.~(\ref{kahlerdec}), with $R$ a unitary matrix. Of course
physical quantities will not depend on $R$ and for any choice we must always
obtain the same physical result. This is due to the invariance of the
Lagrangian under the so-called Weak Basis Transformations (WBT) \cite%
{WBT,quico}. The theory is invariant if we transform the fields as, 
\begin{equation*}
q_{L}=R_{q}q_{L}^{\prime }\quad ;\quad u_{R}=R_{u}u_{R}^{\prime }\quad
;\quad d_{R}=R_{d}d_{R}^{\prime }
\end{equation*}%
where $R_{q}$, $R_{u}$ and $R_{d}$ are transformations from the global
unitary groups $U(3)_{L}$, $U(3)_{u_{R}}$ and $U(3)_{d_{R}}$ respectively,
while simultaneously the Yukawa couplings are transformed as, 
\begin{equation}
Y_{u}^{\prime }=R_{q}^{\dagger }Y_{u}R_{u}\qquad Y_{d}^{\prime
}=R_{q}^{\dagger }Y_{d}R_{d}  \label{yukawa:wbt}
\end{equation}%
Therefore when we choose the different $R_{a}$ all we are doing is to choose
a particular weak basis where we write our theory and the physical results
are absolutely independent of this choice. However, it is very useful to
choose the unitary transformation $R$ in the definition of $N=K^{-1/2}$ to
get a simpler form for this transformation. The form that proves to be
useful is the Cholesky decomposition of an Hermitian matrix. It is always
possible to write an Hermitian matrix as $K=U^{\dagger }U$ in terms of an
upper $U$ triangular matrix, 
\begin{equation}
K=%
\begin{pmatrix}
K_{11} & K_{12} & K_{13} \\ 
K_{12}^{\ast } & K_{22} & K_{23} \\ 
K_{13}^{\ast } & K_{23}^{\ast } & K_{33}%
\end{pmatrix}%
=U^{\dagger }U=%
\begin{pmatrix}
u_{11} & 0 & 0 \\ 
u_{12}^{\ast } & u_{22} & 0 \\ 
u_{13}^{\ast } & u_{23}^{\ast } & u_{33}%
\end{pmatrix}%
\begin{pmatrix}
u_{11} & u_{12} & u_{13} \\ 
0 & u_{22} & u_{23} \\ 
0 & 0 & u_{33}%
\end{pmatrix}%
\end{equation}%
This equation is very easy to solve, 
\begin{eqnarray}
u_{11}=\sqrt{K_{11}} &u_{12}=\frac{\displaystyle{K_{12}}}{\displaystyle\sqrt{%
K_{11}}}&u_{13}=\frac{K_{13}}{\sqrt{K_{11}}}  \label{utri} \\
u_{22}=\sqrt{K_{22}-\frac{\displaystyle{|K_{12}|^{2}}}{\displaystyle{K_{11}}}%
}~~ &~~u_{23}=\frac{\displaystyle{K_{23}K_{11}-K_{13}K_{12}^{\ast }}}{%
\displaystyle\sqrt{K_{22}K_{11}^{2}-K_{11}|K_{12}|^{2}}}~~&~~u_{33}=\sqrt{%
K_{33}-|u_{23}|^{2}-|u_{13}|^{2}}  \notag
\end{eqnarray}%
The inverse of this upper triangular matrix is also upper triangular, and it
is also easily obtained. Obviously we could have chosen to use lower
triangular matrices $L$ instead of the upper triangular matrices $U$ and the
explicit form of the $L$ would then have been obtained in a similar way in
terms of $K$.

This form for the square root of the K\"{a}hler matrix is different from the
\textquotedblleft standard\textquotedblright\ form used in the literature 
\cite{dreiner,peddie}. Clearly the \textquotedblleft
standard\textquotedblright\ form is related to our triangular form by an
unobservable WBT and therefore the two forms are physically
indistinguishable. However it is evident from Eq.~(\ref{utri}) that from the
point of view of calculability it is much simpler to obtain the triangular
form than the \textquotedblleft standard\textquotedblright\ form.

\section{The K\"ahler corrections to Yukawa couplings}

\subsection{The form of the Yukawa coupling matrix}

To proceed we need to know the form of the Yukawa couplings coming from the
superpotential. A fit to the data using a form for the Yukawa matrices where
the smallness of CKM mixing angles is due to the smallness of the
off-diagonal entries with respect to the relevant diagonal entry yields the
structure \cite{romanino}, 
\begin{equation*}
Y_{d}\propto \left( 
\begin{array}{ccc}
0 & \bar{\varepsilon}^{3} & {\ \bar{\varepsilon}^{3}} \\ 
. & {\bar{\varepsilon}^{2}} & {\ \bar{\varepsilon}^{2}} \\ 
. & . & 1%
\end{array}%
\right) ,~~~~~~~~~~Y_{u}\propto \left( 
\begin{array}{ccc}
0 & {\ \varepsilon ^{3}} & {\ \varepsilon ^{3}} \\ 
. & {\ \varepsilon ^{2}} & \varepsilon ^{2} \\ 
. & . & 1%
\end{array}%
\right)
\end{equation*}%
with the expansion parameters $\bar{\varepsilon}=0.15$ and $\varepsilon
=0.05 $. Some non-Abelian family symmetry models can provide such a
structure quite naturally \cite{kingross,spontaneous}. Here we have
suppressed coefficients of ${\mathcal{O}}(1)$. This structure has $%
Y_{kj}<Y_{ij}$ for $i>k$ and $\ j\geq i$ and is unique if the contribution
to the left-handed mixing angles from the elements below the diagonal are
negligible. If one relaxes this constraint then it is possible for some of
the entries to be smaller or zero (texture zeros). We will discuss both
these possibilities. To do so let us first note that, although there are no
direct bounds on the Yukawa couplings below the diagonal from (right-handed)
mixing angles, we can obtain some upper bounds on these entries from their
contributions to the mass eigenvalues. Just requiring that the determinant
of the down Yukawa matrix is $\bar{\varepsilon}^{6}=1\times \frac{m_{s}}{%
m_{b}}\times \frac{m_{d}}{m_{b}}$ we arrive to the conclusion that $%
Y_{21}^{d}\leq \bar{\varepsilon}^{3}$, $Y_{31}^{d}\leq \bar{\varepsilon}$
and $Y_{32}^{d}\leq 1$, assuming no cancellation between different
contributions to the determinant. With this the most general
hierarchical down-quark Yukawa structure consistent with the masses and 
mixing angles is 
\begin{equation}
Y_d\propto \left( 
\begin{array}{ccc}
\leq \bar{\varepsilon} ^{4} & a\text{ }\bar{\varepsilon} ^{3} & {\ b}\text{ }{\bar{\varepsilon} ^{3}}
\\ 
\leq \bar{\varepsilon} ^{3} & c\text{ }{\bar{\varepsilon} ^{2}} & {\ d}\text{ }{\bar{\varepsilon} ^{2}}
\\ 
\leq \bar{\varepsilon} & \leq 1 & 1%
\end{array}%
\right).  \label{genform}
\end{equation}%
Not all of the four coefficients $a,b,c,d$ must be ${\mathcal{O}}(1)$
allowing for the possibility of additional texture zeros. In principle
the $Y_{u}$ structure could also be described by this structure with the only
replacement $\bar{\varepsilon} \to \varepsilon$.  As explained 
below, given that the SM gauge group does not relate the up and down 
right handed sectors, this structure with different expansion parameters
in $Y^u$ and $Y^d$ emerges naturally in a multitude of flavour models both with
Abelian and non-Abelian symmetries, for example in a $U(1)$ model with
Frogatt-Nielsen messenger fields of different masses \cite{ibanez-ross}. 
However, our results below do not require the presence of two 
different 
expansion parameters for the up and the down sector and we could reproduce 
the same fit with different powers of the same expansion parameter
\cite{ramond}\footnote{Strictly speaking the observed values of up-quark 
masses and
CKM mixing angles would still allow $(Y_u)_{23} = {\mathcal{O}}(\varepsilon)$  
and/or $(Y_u)_{13} = {\mathcal{O}}(\varepsilon^2)$ if simultaneously 
$(Y_u)_{32} = {\mathcal{O}}(\varepsilon)$ and 
$(Y_u)_{31} = {\mathcal{O}}(\varepsilon^2)$. Given that this structure
is hierarchical, all the results presented in the following are also valid in  
this case.}.

In this paper we consider the case that this hierarchical structure 
\eq{genform} is reproduced by the
terms of the superpotential in the symmetry basis and we show that the
effect of the K\"ahler potential is then always subdominant in its effects
on the masses and mixing angles.

\subsection{The K\"ahler corrections}

It proves to be useful in most realistic models to go to the canonically
normalised basis by redefining the fields by a wave function normalisation
matrix chosen to have the upper triangular form, as discussed above. Using
this form the correction to the Yukawa coupling matrix in the Standard Model
(SM) is of the form 
\begin{equation*}
H \hat {\overline{\psi}}_L~ Y~ \hat \psi_R ~~\equiv ~~ H~\hat{\psi}%
_{L\,i}^*~Y_{ij}~\hat{\psi}_{R\,j}~~=~~H~\psi_{L\,k}^*
N_{L\,ik}^*~Y_{ij}~N_{R\,jm}\psi_{R\,m}~~=~~H~\psi_{L\,k}^*~Y_{km}^{t}~
\psi_{R,m}
\end{equation*}%
If we consider, for the moment, only the transformation on the Left Handed 
($LH$) fields using our triangular matrices, with $N=U$, the total (t) 
Yukawa is, 
\begin{eqnarray}
Y_{ij}^{t} &=&\sum_{i\geq k}N_{ki}^{\ast }Y_{kj} \label{totYuk}
\simeq N_{ii}^{\ast }Y_{ij}+\sum_{i>k}N_{ki}^{\ast }Y_{kj}
\end{eqnarray}

As may be seen in Eq.~(\ref{one}) the expansion parameters are given by
terms of the form $<\theta >/M$ where $M$ is the messenger mass. In the
superpotential the expansion parameters come from both the $LH$ and Right 
Handed ($RH$) sectors. The expansion parameters, $\varepsilon $ and 
$\bar{\varepsilon},$
for the up and down sectors\footnote{%
Here we have implicitly assumed that $\epsilon =<\theta >/M$ is the
fundamental expansion parameter. If this is not true and the true expansion
parameter is larger (e.g. $\theta $ is itself generated by a higher
dimension term $\phi .\phi /M)$ one should allow for the possibility that
the expansion parameter in the \textit{K\"ahler }sector is the larger one
(e.g. $<\phi >/M$).} in the superpotential may differ as the SM gauge group
does not relate the up and down right handed quark sectors. However the
contribution from the $LH$ sector to the mass matrix structure must be equal
in the up and down sectors due to the $SU(2)_{L}$ gauge symmetry. Thus its
contribution cannot be larger than $\varepsilon ,$ the smaller of the two
(right handed) expansion parameters. This implies that the K\"ahler
rescaling matrix in the $LH$ sector, $N_{ik}^{L},$ has a strong hierarchy
controlled by the small parameter $\varepsilon $ with $N_{ii}^{L}\simeq 1$
and $N_{ik}^{L}\leq \epsilon $ for the non-zero entries of the upper
triangular form. Notice that the smallness of off-diagonal elements in 
$N_{ij}$  (and $K_{ij}$) is necessary in any model where the hierarchy 
of the Yukawa matrices is due to the presence of an underlying family 
symmetry, either Abelian or non-Abelian. This is specially simple 
in Abelian models where the hierarchy in the Yukawa matrices and CKM mixing 
angles is guaranteed by the different powers of the flavon field. Using a 
triangular form for $N$ and taking into account
that the hierarchy in the left handed angles implies that $q_j > q_i$, we 
have that $N_{i < j} = (\theta^*/M)^{(q_j - q_i)}= \epsilon^{(q_j - q_i)}$. 
This form is forced 
from the requirement of invariance under the Abelian symmetry of the 
canonically normalised Yukawa element. In 
the case of non-Abelian symmetries the hierarchy in the Yukawa matrices is 
obtained from the smallness of vevs of the different flavon fields. 
In principle, only the vev defining the third generation can be 
${\mathcal{O}}(1)$ while vevs in the direction of the second or 
first 
generation are $\leq \epsilon$. Given that the flavour structure of the 
K\"ahler matrices is necessarily generated in terms of the same flavon 
vevs,
we have that any off-diagonal term in the K\"ahler matrices involves at 
least one power of the small vevs and hence $N_{i < j} \leq \epsilon$.      
 A similar argument applies to the up quark 
$RH$ sector, $N_{ii}^{R,u}\simeq 1$ and $N_{ik}^{R,u}\leq \varepsilon$ but in
the down quark $RH$ sector the expansion parameter must be the larger one, $%
\bar{\varepsilon}$, so $N_{ii}^{R,d}\simeq 1$ and $N_{ik}^{R,d}\leq \bar{%
\varepsilon}.$

In fact it is easy to prove that for the hierarchical textures of interest
here the leading correction to a given Yukawa element is suppressed by at
least ${\mathcal{O}}(\epsilon ^{2})$ \footnote{This was first shown in the 
particular case of Abelian flavour symmetries in Ref.~\cite{eyalnir}}. 
With the underlying family symmetry
ordering the correction we know that, before symmetry breaking, the operator
giving rise to the correction to a given element must transform in the same
way under the family symmetry as the leading term. We have just proved that
the difference of the K\"ahler transformations from the identity is at least
of $O(\epsilon )$. Furthermore corrections to $Y_{ij}$ after transformations
to canonical K\"ahler with upper triangular matrices come only from $
Y_{kj}$ with $i>k$ and $Y_{kj}<Y_{ij}$. This implies that a new
contribution to $Y_{ij}^t$ is subdominant relative to $Y_{ij}$ at least
by $O(\epsilon )$ where $\epsilon =<\theta >/M$. As $\theta $ transforms
non-trivially under the family symmetry, to maintain the symmetry property
of the leading term, this relative correction must be given by a combination
of fields which transforms as a singlet, that is at least of the form $%
\theta \theta ^{\dagger }$ and hence of ${\mathcal{O}}(\epsilon ^{2}).$ This
result applies to hierarchical Yukawa structures. For the case that the $%
(2,3)$ element saturates the bound of Eq.~(\ref{genform}) it violates the
condition of hierarchical Yukawa couplings and our conclusions above do not
apply. In what follows we consider this possibility separately.

Using this we will now calculate the canonical Yukawa through Eq.~(\ref%
{totYuk}). Although we have started with the superpotential generating the
form of Eq.~(\ref{genform}) in the symmetry basis we have the freedom to use
any basis when calculating the effects on physical quantities. It is
convenient to go to the Cholesky form when determining the effects of the
K\"ahler potential and we use an upper triangular form for the K\"ahler
rescaling matrix in the $LH$ sector with $N_{ki}^{L}=0$ for $i<k.$ The
corrections to a given element of the Yukawa matrix induced by the
transformation to canonical K\"{a}hler are given by $N_{ki}^*Y_{kj}$.

\subsubsection{No additional texture zeros}

We first consider the case without additional texture zeros so that all of $%
a,b,c,d$ are of ${\mathcal{O}}(1).$ Taking into account that $Y_{kj}<Y_{ij}$
for $i>k$ and $\ j\geq i$ we conclude that $N_{ki}Y_{kj}<Y_{ij}$. Therefore,
these corrections are always sub-dominant in $\epsilon $. This is not yet
sufficient to prove that the transformation to the canonical left handed K%
\"{a}hler basis does not change the observable mixings and masses because
they could be sensitive to elements of $Y$ below the diagonal. Given the
bounds of Eq.~(\ref{genform}) the only terms that could be modified by 
K\"ahler corrections are the $(2,1)$, $(3,1)$ and $(3,2)$ terms.
For instance, for $Y_{3,1}<Y_{2,1}$ the K\"{a}hler correction can dominate 
the $(3,1)$ element. However in this case, from the structure in Eq.~(\ref%
{genform}) and with $N_{i<j} \leq \epsilon$, $Y_{3,1}^{t}\leq \epsilon ^{4}$. 
Clearly this is too small to affect masses or $LH$ mixing angles at 
leading order. It can be easily checked that the same is true in the case of
the $(2,1)$ and $(3,2)$ elements.  As we have discussed,
for the hierarchical textures of interest here, the leading correction to a
given Yukawa element is suppressed by at least ${\mathcal{O}}(\epsilon ^{2})$.

One might worry that the condition $Y_{kj}<Y_{ij}$ for $i>k$ and $\ j\geq i$
is too strong and that what are constrained are the elements after K\"{a}%
hler mixing, i.e. $Y_{kj}^{t}<Y_{ij}^{t}$ for $i>k$ and $\ j\geq i$ and the
condition on $Y_{kj}$ is not satisfied. However this is inconsistent. To see
this note that the phenomenological structure of $Y_{kj}^{t}$ in Eq.~(\ref%
{genform}) would correspond both to the basis of canonical K\"{a}hler with
upper triangular transformations or to the basis of "standard" canonical
transformations. This is due to the fact that both basis are related by a
small rotation which does not change the order of the elements if the
departure of the original K\"{a}hler metric from the identity is also
hierarchical as expected in models with a spontaneously broken family
symmetry. Thus we still have $N_{ik}\leq \epsilon $ for $i\neq k$.
Therefore, we would need $Y_{kj}>Y_{ij}$ for $i>k$, or more exactly the
power in $\epsilon $ of $Y_{kj}$ is smaller than the power in $\epsilon $ of 
$Y_{ij}$ for $i>k$ so that $N_{ki}^{\ast }Y_{kj}>Y_{ij}$ is possible.
However in this case we necessarily have $%
Y_{kj}^{t}=Y_{kj}>Y_{ij}+N_{ik}Y_{kj}=Y_{ij}^{t}$ for $i>k$ and $\ j\geq i$
(neglecting smaller contributions from $Y_{mj}$ with $m<k$ if present) and
we arrive to an inconsistency with the initial statement $%
Y_{kj}^{t}<Y_{ij}^{t}$. Thus even with the weaker condition we need $%
Y_{kj}<Y_{ij}$ for $i>k$ and $\ j\geq i.$

So far we have discussed the transformations to canonical K\"{a}hler for the
left handed fields. Now, we have to proceed exactly in the same way for the
right-handed transformation. Clearly, if the Yukawa structures are also
hierarchical we can perform the same analysis using upper triangular
matrices and we would again arrive to the conclusion that corrections from
the K\"{a}hler to any Yukawa element are always sub-dominant at least by $%
\epsilon ^{2}$ ($\epsilon =\bar{\varepsilon},\varepsilon $ for $%
Y=Y_{d},Y_{u}).$ There is an exception to this conclusion if $Y_{23}$ does
not preserve the hierarchical structure and is of ${\mathcal{O}}(1)$
saturating the bound in Eq.~(\ref{genform}). In this case it is possible
that $N_{23}^{R}={\mathcal{O}}(1)$ and therefore corrections ${\mathcal{O}}%
(1)$ to $Y_{i3}$ are still possible. Even in this case, it is clear that we
can never modify the order in $\epsilon $ of the different elements of the
Yukawa matrix, all it can do is to change the ${\mathcal{O}}(1)$
coefficients of the $Y_{i3}$ elements. To determine whether this special
case is possible one needs to know $Y_{32}$ and this can be done through
measurement of flavour changing neutral currents \cite{FCNC,b2s} or lepton
flavour violation \cite{LFV}.

Thus, using the triangular form, we have shown that the K\"{a}hler
corrections to the Yukawa matrix are sub-dominant for hierarchical Yukawa
matrices. In the next section we prove that this is also true for the
observable mixing angles and mass eigenstates.

\subsubsection{ Additional texture zeros}

A special situation occurs when one of $a,b,c,d$ is $<{\mathcal{O}}(1)$
giving rise to an approximate texture zero. This can spoil the hierarchical
structure of our Yukawa textures, $Y_{kj}<Y_{ij}$ for $i>k$ and $\ j\geq i$
and therefore must be analysed separately. An example of the origin of such
zeros occurs in spontaneously broken Abelian theories through the so-called
holomorphic zeros \cite{dudas}. In this case the symmetry breaking is
through flavon field(s) carrying only one sign of charge (say negative) and
then a net negative charge of the fermionic fields cannot be compensated
with insertions of the flavon field because, due to the holomorphicity of
the superpotential, the charged conjugated flavon can not be used. However
the K\"{a}hler potential is non holomorphic and therefore these zeros can be
filled after the transformation to the canonical basis.

As before, if we are only interested in the physical effects of this texture
zero filling we can choose a convenient basis \cite{dudas}. Once more our
choice of upper triangular matrices is especially simple. In a hierarchical
texture we can have a texture zero in any position of the matrix except in $%
Y_{33}$ which is necessarily ${\mathcal{O}}(1).$ Although it is clear that
the texture zeros can be filled in by the K\"ahler corrections we can
immediately use the analysis presented above to show that physical
measureables will not be affected by these corrections. The point, as is
explicitly demonstrated in the next section, is that the form of Eq.~(\ref%
{genform}) gives the value of each entry of the Yukawa matrix that has a
leading effect on a mass or a mixing angle. If the entry is larger than the
value shown it will give a mass or mixing angle in conflict with the
measured value. If the entry is smaller it will only contribute to
measureable quantities at subleading order.

In the previous section we showed that, for the case of hierarchical
textures, the K\"ahler corrections only contribute to the Yukawa matrix
elements suppressed relative to the order shown in Eq.~(\ref{genform}) by at
least ${\mathcal{O}}(\epsilon ^{2}).$ For example we can see that a zero in $%
Y_{11}$ is never filled by any other element. In the same way a zero in $%
Y_{12}$ or $Y_{21}$ is only filled by a non-zero entry in $Y_{11}$. Taking
into account the constraints from the determinant of the Yukawa matrix, $%
Y_{11}\leq \epsilon ^{4}$ and in the hierarchical case with $N_{i\neq
j}^{L(R)}\leq \epsilon $ this implies that they can only be filled at ${%
\mathcal{O}}(\epsilon ^{5})$. In the same way $Y_{13}$ $Y_{31}$ and $Y_{22}$
can only be filled at ${\mathcal{O}}(\epsilon ^{4})$ ($Y_{12},Y_{21}\leq
\epsilon ^{3}$). Finally a zero in $Y_{23}$ or $Y_{32}$ implies that $%
Y_{22}=\epsilon ^{2}$ and hence these zeros can be filled at most at ${%
\mathcal{O}}(\epsilon ^{3})$. As we will now show, these subleading terms
only contribute to physical quantities at subleading order even though the
texture zero may be filled in. The only exception to this is when the
hierarchical structure is spoilt through an ${\mathcal{O}}(1)$ term in $%
Y_{23}.$ In this case, following the discussion given above, the K\"ahler
corrections can contribute at ${\mathcal{O}}(1)$ to physical quantities.

\section{K\"ahler corrections to the mass matrix eigenvalues and mixing
angles}

To complete our proof we need to demonstrate that the entries of Eq.~(\ref%
{genform}) are the smallest that can affect masses and mixing angles and
thus the K\"ahler corrections, which we have shown are smaller than those of
Eq.~(\ref{genform}), are necessarily subdominant in determining physical
quantities.

\subsection{Quark and charged lepton masses and mixing angles.}

Since the K\"{a}hler corrections are wave function corrections which cannot
change the rank of the mass matrix we know that they lead to multiplicative
normalisations of the masses. For hierarchical Yukawa matrices the wave
function normalisation has the form $N_{ik}=\delta _{ik}+{\mathcal{O}}(\leq
\varepsilon ) $ and this means the K\"{a}hler corrections to masses are
necessarily sub-dominant. To see this explicitly, consider only the left
handed canonical normalisation $N_{ik}$ with $N$ upper triangular. Now using
Eq.~(\ref{totYuk}) the canonical Yukawa and the fact that the Yukawa and K%
\"{a}hler matrices are hierarchical in the left handed sector, the
determinant of $Y^{t}$ is, 
\begin{equation*}
\mbox{Det}(Y^{t})=\mbox{Det}(N)\mbox{Det}(\hat{Y})\simeq (1+{\mathcal{O}}%
(\leq \epsilon ))\mbox{Det}(\hat{Y})
\end{equation*}%
Moreover, from the hierarchical structure in Eq.~(\ref{genform}) we know
that any element of the matrix is corrected only at ${\mathcal{O}}(\leq
\epsilon ^{2})$ under the transformations to canonical left-handed K\"{a}%
hler. In particular, the heaviest eigenvalue in $Y^{t}$ will be still be $1+{%
\mathcal{O}}(\leq \epsilon ^{2})$. Therefore this implies that the product
of the two lightest eigenvalues can only be changed at ${\mathcal{O}}(\leq
\epsilon ^{2})$. Finally the second eigenvalue is basically obtained from
the lightest eigenvalue of the $(2,3)$ submatrix and thus we obtain again
that any change to this eigenvalue will be sub-dominant in $\epsilon $ and
therefore the same is true for the first generation eigenvalue.

In the case of a non-hierachical structure in the $(2,3)$ entry with $%
N_{23}^{R}$ of ${\mathcal{O}}(1)$ we expect $\mbox{Det}(N^{R})$ to be ${%
\mathcal{O}}(1)$ barring accidental cancellations. In this case the
corrections to the eigenvalues, while still not changing their order in $%
\epsilon ,$ could be ${\mathcal{O}}(1)$.

Concerning the mixing angles, with the use of triangular matrices we have
not changed the hierarchical structure of the Yukawa matrices. Hence, we can
still use the usual perturbative expansion. In this way, after the
transformations to left handed canonical K\"ahler we have, 
\begin{eqnarray}
\theta_{23} &=& \theta_{23}^d - \theta_{23}^u = \frac{(Y^{d}_{23})^t}{%
(Y^{d}_{33})^t} - \frac{(Y^{u}_{23})^t}{(Y^{u}_{33})^t} = \frac{\hat
Y^d_{23} ( 1 +{\mathcal{O}}( \varepsilon^2))}{ \hat Y^d_{33}( 1 +{\mathcal{O}%
}(\varepsilon^2)) } - \frac{ \hat Y^u_{23}( 1 +{\mathcal{O}}(\varepsilon^2))%
}{\hat Y^u_{33}( 1 +{\mathcal{\ O}}(\varepsilon^2))} \notag \\ &=& \hat \theta_{23} ( 1 +%
{\mathcal{O}}(\varepsilon^2))
\end{eqnarray}
the discussion is identical for the $\theta_{13}$ mixing angle. The case of $%
\theta_{12}$ is slightly more complicated, now we have, 
\begin{eqnarray}
\theta_{12}^d = \frac{\displaystyle{(Y_{12}^{d})^t}} {\displaystyle{%
(Y_{22}^{d})^t -(Y_{23}^{d})^t (Y_{32}^{d})^t}}
\end{eqnarray}
where the denominator is really the $Y_{22}^d$ element in the basis where we
have already diagonalised the $2,3$ sector, and it is approximately equal to 
$m_s/m_b = \bar \varepsilon^2$. However, we know that both $(Y_{22}^d)^t
\leq \bar \varepsilon^2 ( 1 +{\mathcal{O}}(\varepsilon^2))$ and $%
(Y_{23}^d)^t (Y_{32}^d)^t \leq \bar \varepsilon^2 ( 1 +{\mathcal{O}}%
(\varepsilon^2))$. This means that the denominator can also be corrected
only at ${\mathcal{O}}(\varepsilon^2)$, then we have, 
\begin{eqnarray}
\theta_{12}^d = \frac{\displaystyle{\hat Y_{12}^d (1 +{\mathcal{O}}%
(\varepsilon^2))}}{\displaystyle{(\hat Y_{22}^d - \hat Y_{23}^d \hat
Y_{32}^d)( 1 +{\mathcal{O}}(\varepsilon^2)) }} = \hat \theta_{12}^d ( 1 +{%
\mathcal{O}}(\varepsilon^2))
\end{eqnarray}
doing the same for $\theta_{12}^u$ we arrive immediately to $\theta_{12} =
\theta_{12}^d - \theta_{12}^u = \hat \theta_{12} ( 1 +{\mathcal{O}}%
(\varepsilon^2))$.

Moreover, it is easy to check that the effect of the transformation to
canonical K\"{a}hler for the right handed fields on the left handed mixings
is usually negligible. To see this, we consider the limit of trivial left
handed K\"{a}hler and nontrivial right-handed K\"{a}hler. Then, we consider
the diagonalisation of the Hermitian matrix $H^{t}$, 
\begin{equation*}
H^{t}=Y^{t}(Y^{t})^{\dagger }=\hat{Y}N_{R}N_{R}^{\dagger }\hat{Y}^{\dagger }=%
\hat{Y}K^{-1}\hat{Y}^{\dagger }=\hat{V}_{L}^{\dagger }\hat{M}_{f}\hat{V}%
_{R}K^{-1}\hat{V}_{R}^{\dagger }\hat{M}_{f}V_{L}\equiv \hat{V}_{L}^{\dagger }%
\hat{M}_{f}\tilde{K}^{-1}\hat{M}_{f}\hat{V}_{L}
\end{equation*}%
where we have written $\hat{Y}=\hat{V}_{L}^{\dagger }\hat{M}_{f}\hat{V}_{R}$
and reabsorbed the right-handed rotation in $\tilde{K}^{-1}$, i.e. we have
written the inverse of the K\"{a}hler in the basis of right handed mass
eigenstates. Now it is trivial to see that the matrix diagonalising $H^{t}$
will be the product of $V_{L}$ with the matrix diagonalising $\hat{M}_{f}%
\tilde{K}^{-1}\hat{M}_{f}$. As we have seen $\hat{M}_{f}$ are approximately
equal to the eigenvalues of the total Yukawa matrix, this implies that $\hat{%
M}_{f}\tilde{K}^{-1}\hat{M}_{f}$ is strongly hierarchical and then the
mixing angles diagonalising this matrix will be, 
\begin{equation*}
\tilde{\theta}_{i3}\simeq \frac{\displaystyle{m_{i}m_{3}(\tilde{K}^{-1})_{i3}%
}}{\displaystyle{m_{3}^{2}(\tilde{K}^{-1})_{33}}}\qquad \tilde{\theta}%
_{12}\simeq \frac{\displaystyle{m_{1}m_{2}(\tilde{K}^{-1})_{12}}}{%
\displaystyle{m_{2}^{2}\left( (\tilde{K}^{-1})_{22}-\frac{|(\tilde{K}%
^{-1})_{23}|^{2}}{(\tilde{K}^{-1})_{33}}\right) }}
\end{equation*}%
therefore these contributions are suppressed both by the smallness of
off-diagonal entries in the K\"{a}hler with respect to diagonal ones and by
ratios of fermion masses. This last suppression is usually enough to make $%
\tilde{\theta}_{ij}\ll \theta _{ij}$ and then we can safely neglect the
effect of right handed transformation in left handed mixings.

The exception to this rule arises when the right handed K\"{a}hler in the
basis of right handed mass eigenstates is not hierarchical and has ${%
\mathcal{O}}(1)$ entries in $K_{23}$, $K_{22}$ and $K_{33}$. In this case
the correction to the angle $\theta _{23}$ from the down quark right handed K%
\"{a}hler could be of leading order as both $\theta _{23}$ and $m_{s}/m_{b}$
are ${\mathcal{O}}(\bar{\varepsilon}^{2})$. Still this situation can be
understood as an exception to the main rule we formulated above. The
correction from the right handed K\"{a}hler in the left handed mixing angles
would still be of the same order as the contribution from the non-canonical
Yukawa matrix and therefore would only modify the unknown ${\mathcal{O}}(1)$
coefficients. Usually, we find this structure in $U(1)$ models with lopsided
Yukawa textures \cite{lopsided}. These models depend precisely on the
existence of different ${\mathcal{O}}(1)$ coefficients in the elements of
the Yukawa texture to obtain the correct masses and mixing angles. However,
the $U(1)$ symmetry has no control on these ${\mathcal{O}}(1)$ coefficients
and so this means that we do not need to worry about these effects. Only in
a theory where we can control these unknown coefficients we should worry
about the effects of this right-handed field redefinition.

\subsection{Neutrino masses and mixing angles}

The case of neutrino masses can be analysed with similar techniques. In this
case, we obtain the effective Majorana mass matrix for the left handed
neutrinos through the seesaw mechanism.\ The neutrino mass matrix structure
has the form%
\begin{equation*}
L_{\nu }=-\nu _{L\,i}Y_{ij}^{\nu }\nu _{R\,j}^{c}-\frac{1}{2}\nu
_{R\,i}M_{R\,ij}\nu _{R\,j}^{c}+h.c.
\end{equation*}%
giving the effective Majorana mass matrix of the effective low energy
neutrinos, $M_{\nu }$ of the form 
\begin{equation*}
M_{\nu }=\chi _{\nu }~(v\sin \beta )^{2}=Y^{\nu }(M_{R})^{-1}Y^{\nu
\,T}~~(v\sin \beta )^{2}
\end{equation*}
The transformation properties of the effective neutrino mass matrix under
the transformations to canonical K\"{a}hler\ for both left handed and right
handed fields is given by 
\begin{eqnarray}
\chi _{\nu }^{t} &=&Y^{\nu \,t}(M_{R}^{t})^{-1}(Y^{\nu \,t})^{T}=N_{L}^{T}~%
\hat{Y}^{\nu }~N_{R}(N_{R})^{-1}~M_{R}^{-1}~(N_{R}^{T})^{-1}N_{R}^{T}~\hat{Y}%
^{\nu \,T}~N_{L}  \notag \\
&=&~~N_{L}^{T}~\hat{Y}^{\nu }~\hat{M}_{R}^{-1}~\hat{Y}^{\nu
\,T}~N_{L}~~=~~N_{L}^{T}~\hat{\chi}_{\nu }~N_{L}
\end{eqnarray}%
Hence, we see that the effective neutrino coupling $\chi _{\nu }$ is
transformed only by the left handed canonical transformations and the
right-handed transformations cancel exactly.

However the neutrino sector can be special because in this case, we do not
know much about the hierarchy of the leptonic Yukawa couplings $Y^\nu$ and $%
Y^e$. In fact we can find two different situations:

\begin{enumerate}
\item $Y^{\nu }$ and $Y^{e}$ are hierarchical and $Y_{kj}<Y_{ij}$ for $i>k$
and $\ j\geq i$. This is this situation in realistic non-Abelian flavour
theories explored to date \cite{kingross}.

\item $Y^\nu$ or $Y^e$ have two rows of similar size. We can find this
situation in some $U(1)$ models \cite{King:1999cm}.
\end{enumerate}

In case 1 the K\"{a}hler metric is also very close to the identity with
small off-diagonal entries. Therefore we can choose $N_{L}$ to be upper
triangular with $(N_{L})_{ii}\simeq 1$ and $(N_{L})_{ij}\leq \epsilon $.
Then both $Y_{\nu }$ and $Y_{e}$ are only changed at higher order in $%
\epsilon $ and neutrino masses and mixings are only changed at sub-dominant
order. In the case of non-Abelian symmetries $\chi _{\nu }^{t}$ and $Y^{e}$
are changed at most at order $\varepsilon ^{2}$. Then we can immediately use
the standard formulae for the neutrino mixings compiled in Ref. \cite{steve}%
. For all the different cases compatible with hierarchical rows in the
lepton Yukawa matrix, we can immediately see that neutrino mixings will only
be changed at sub-leading order. Although small, this might still be
relevant for the difference of the solar mixing angle from maximality 
\cite{nirjhep}.

Case 2 arises if two left handed fields have identical flavour symmetry
charges. As a result the K\"{a}hler metric will have large mixing between
these two fields and therefore ${\mathcal{O}}(1)$ off-diagonal entries. In
this case, it is possible to modify the ${\mathcal{O}}(1)$ coefficients in
the different elements of the canonical Yukawa matrices, but the order in $%
\epsilon $ of these entries is not changed. Therefore, in this case, it is
possible to generate changes at leading order in neutrino masses and
mixings. This corresponds again to the case where right-handed mixing angles
can modify left-handed mixings in the quark sector. Since only the ${%
\mathcal{O}}(1)$ coefficients are modified these corrections do not change
the predicted structure if the family symmetry does not predict the value of
these coefficients.

\subsection{Soft SUSY breaking masses and mixing angles}

Finally, we would also like to comment on the effects of the K\"{a}hler
transformations on the soft breaking masses which may give rise to dangerous
flavour changing neutral current processes \cite{FCNC}. Notice that the
F-term contributions to soft breaking masses in supergravity are closely
related to the K\"{a}hler potential \cite{soni}. In fact the non canonical
soft breaking masses are, 
\begin{equation*}
\hat{m}_{\overline{a}b}^{2}=m_{3/2}^{2}{K}_{\overline{a}b}-F_{\overline{m}%
}\left( \partial _{\overline{m}}\partial _{n}{K}_{\overline{a}b}-\partial _{%
\overline{m}}{K}_{\overline{a}c}({K}^{-1})_{c\overline{d}}\partial _{n}{K}_{%
\overline{d}b}\right) F_{n}
\end{equation*}%
To obtain the canonical soft breaking masses we have to multiply this matrix
by the inverse of the square root of $K$, $m^{2}=(K^{-1/2})^{\dagger }\hat{m}%
^{2}K^{-1/2}$. Then we obtain, 
\begin{eqnarray*}
m^{2} &=&m_{3/2}^{2}\mathbf{1}-(K^{-1/2})^{\dagger }F_{\overline{m}}\left(
\partial _{\overline{m}}\partial _{n}{K}-\partial _{\overline{m}}{K}({K}%
^{-1})\partial _{n}{K}\right) F_{n}K^{-1/2} \\
&\equiv &m_{3/2}^{2}\mathbf{1}-N^{\dagger }F_{\overline{m}}\left( \partial _{%
\overline{m}}\partial _{n}{K}-\partial _{\overline{m}}{K}({K}^{-1})\partial
_{n}{K}\right) F_{n}N
\end{eqnarray*}%
Therefore we see that we have a universal contribution proportional to $%
m_{3/2}^2$ plus other terms which in principle will depend on flavour. These
terms depend on the derivatives of the K\"ahler potential with respect to
fields with non vanishing F-terms.

If the field with non-vanishing F-term is a hidden sector field it must be
neutral under the flavour symmetry and therefore the structure in powers of $%
\epsilon$ of $\partial_{\overline{m}}\partial_n {K}$ or $\partial_{\overline{%
m}} {K}$ will be the same as the structure of $K$. However, factors ${%
\mathcal{O}}(1)$ can be different and indeed can sometimes be zero. The
important point is that no terms larger in powers of $\epsilon$ are
generated than are in $K$ itself. Due to this difference in the ${\mathcal{O}%
}(1)$ coefficients the product $(K^{-1/2})^\dagger\partial_{\overline{m}}{K}
K^{-1/2}$ will be different from the identity, but will be bounded by the
same power in $\epsilon$ as the original $K$ matrix \cite{timjones}.

Another possibility is that the field with non-vanishing F-term is a flavon
field with non-trivial quantum numbers under the flavour symmetry. As shown
in \cite{Fterms}, the natural size for $F_\theta$ for $\theta$ a flavon
field is $m_{3/2} \langle \theta \rangle$, although it can be smaller
depending on the characteristics of the scalar potential. In this case, we
also have that $F_{\overline{m}} \partial_{\overline{m}} {K}$ cannot
generate terms larger in powers of $\epsilon$ than the terms initially
present in $K$ itself and the conclusion above still applies.

We have also to consider the possibility of a non-vanishing flavour D-term
contributing to the soft masses. Although this possibility is extremely
dangerous for the phenomenology of flavour changing neutral currents (FCNCs)
it can be realised for heavy sfermion masses in some Abelian flavour models.
In this case we obtain a new contribution to the soft masses, 
\begin{equation*}
(\hat{m}_{\overline{a}b}^{2})^{D}~=~g~q_{b}~K_{\overline{a}b}~\langle
D\rangle
\end{equation*}
with $q_{b}$ the charge of the field $\phi _{b}$ under the $U(1)_{fl}$
symmetry. Notice that due to the dependence on the charges of the different
fields this contribution to the soft masses is not diagonalised when we make
the transformation to the basis of canonical K\"{a}hler and therefore it
gives rise to new FCNC effects.

To analyse these FCNC effects it is convenient to work in the SCKM basis
where the corresponding Yukawa matrix is diagonal. Therefore, to obtain the
sfermion mass matrix in the SCKM basis we have to do two trasformations.
First we go to the basis of canonical K\"ahler with our triangular matrices
and second we diagonalise the corresponding Yukawa matrix with a rotation of
the full superfield. Now, we can compare the effects of the transformations
to the basis of canonical K\"{a}hler with the effects of the second
transformation to the SCKM basis. First it is easy to see that in $U(1)$
models the structure in $\epsilon $ of our triangular K\"{a}hler
transformations are always smaller or equal that the corresponding rotation
diagonalising the Yukawa matrix. For instance, the left handed K\"{a}hler
transformation is usually of the same order as the left handed rotation
diagonalising the up quark Yukawa matrix and smaller than the left handed
rotation diagonalising the down quark Yukawa. If the diagonal elements of
the K\"{a}hler metric are ${\mathcal{O}}(1)$, this means that the
corrections to offdiagonal elements that we obtain from the transformations
to the SCKM basis are larger or equal than the corrections obtained in the
transformation to the canonical basis. As before, if we are not interested
in coefficients ${\mathcal{O}}(1)$, we can also ignore the effects of
transformation to canonical K\"{a}hler in the soft breaking masses.

\section{Conclusions}

In this letter we have studied the effects of the transformations to the
canonical K\"{a}hler basis on the Yukawa textures for quarks and leptons and
their contributions to physical masses and mixing angles. We have developed
a simple formalism that allows a straightforward calculation of the
necessary K\"{a}hler transformations and simplifies enormously the
phenomenological analysis. Using this formalism we have proved that, in the
case of models with a hierarchical structure of the Yukawa matrices, the
corrections obtained through the transformations to canonical K\"{a}hler are
always suppressed by a factor $\leq \epsilon ^{2}$ with $\epsilon $ the
expansion parameter in the Yukawa matrix. This implies that, in this case,
fermionic masses and mixing angles receive only corrections at $\epsilon
^{2} $ from the K\"{a}hler transformations. We have seen that although
texture zeros can be filled by transformations to canonical K\"{a}hler the
physical effects of this texture zero filling are only subdominant
corrections in $\epsilon $ to observable masses and mixing angles. We have
also discussed some exceptions to the case of completely hierarchical Yukawa
matrices where some corrections at leading order are possible. In any case,
we have seen that in these models only unknown ${\mathcal{O}}(1)$
coefficients are modified. We have also shown that the corrections to the
scalar soft breaking mass matrices can only change the unknown ${\mathcal{O}}%
(1)$ coefficients. We conclude that in the large class of models considered
here the leading order superpotential couplings in the noncanonical K\"{a}%
hler basis are essentially unchanged when transformed to the canonical K%
\"{a}hler basis.

\section*{Acknowledgements}

L. V-S. would like to thank G. Kane for helpful discussions. O.V.
acknowledges partial support from the spanish MCYT FPA2002-00612 and Andrea
Romanino for useful discusions.

\section*{Note Added}

During completion of this work, we learnt about two groups \cite%
{alejandro,tim} working along the same lines with different techniques. Our
conclusions agree in the points where the analysis overlap. In Ref. \cite%
{alejandro}, the authors provide an exact formula relating the
\textquotedblleft naive\textquotedblright\ CKM and MNS matrices to the
physical matrices. They show that the effects of canonical normalisation are
subdominant in the case of hierarchical matrices in agreement with the
present analysis which uses somewhat simpler mathematical techniques to
obtain the transformation to the canonical K\"{a}hler basis. In addition we
analyse the effects of canonical normalisation in the sfermion mass matrices
and we find that these transformations do not change the structure of the
sfermion mass matrices.

\end{document}